\documentclass{emulateapj}
\pdfoutput=1
\usepackage{apjfonts}
\usepackage{epsfig}
\usepackage{natbib}
\def\gtrsim{\mathrel{\hbox{\rlap{\hbox{\lower4pt\hbox{$\sim$}}}\hbox{\raise2pt\hbox{$>$}}}}}

\newcommand{\mbh}{\ensuremath{M_\mathrm{BH}}}

\newcommand{\msun}{\ensuremath{M_{\odot}}}

\def\lax{{$\mathrel{\hbox{\rlap{\hbox{\lower4pt\hbox{$\sim$}}}\hbox{$<$}}}$}}
\def\gax{{$\mathrel{\hbox{\rlap{\hbox{\lower4pt\hbox{$\sim$}}}\hbox{$>$}}}$}}

\slugcomment{To appear in {\it Nature Communications}.}
\shorttitle{{\it Seed Black Holes}}
\shortauthors{GREENE}

\begin{document}

\title{Low-mass black holes as the remnants of primordial black hole 
formation}

\author{Jenny E. Greene}
\affil{Department of Astrophysical Sciences, Princeton University, Princeton, 
NJ 08544}

\section{Abstract}

This article documents our ongoing search for the elusive
``intermediate-mass'' black holes. These would bridge the gap between
the approximately ten solar mass  (\msun) ``stellar-mass'' black holes (the
end-product of the life of a massive star) and the ``supermassive''
black holes with masses of millions to billions of solar masses found
at the centers of massive galaxies.  The discovery of black holes with
intermediate mass is the key to understanding whether supermassive
black holes can grow from stellar-mass black holes, or whether a more
exotic process accelerated their growth only hundreds of millions of
years after the Big Bang.  Here we focus on searches for black holes
with \mbh$\sim 10^4-10^6$ solar masses that are found at galaxy
centers.  We will refer to black holes in this mass range as
``low-mass'' black holes, since they are at the low-mass end of
supermassive black holes. We review the searches for low-mass black
holes to date and show tentative evidence, from the number of low-mass
black holes that are discovered today in small galaxies, that the
progenitors of supermassive black holes were formed as ten thousand to
one-hundred thousand solar mass black holes via the direct collapse of
gas.

\section{The Black Hole Mass Scale}

Over the last decade we have come to understand that supermassive
black holes, with masses of millions to billions of times the mass of
the Sun, are very common in the centers of massive galaxies
\citep{richstoneetal1998}.  Black holes are found in the centers of 
most massive galaxies at the present time.  We would like to 
understand when and how they formed and grew.

We cannot yet watch the first supermassive black holes form. They
did so soon after the Big Bang, and light from those distant events is
beyond the reach of today's telescopes.  However, we do have two very
interesting limits on the formation of the first black holes.  The
first comes from observations of the most distant known
black holes: light is emitted by material falling into the deep gravitational 
potential of the black hole. These monsters are so bright that
they must be powered by at least billion solar mass black holes.
They had very little time to grow, as we see them only a few hundred
million years after the Big Bang \citep{fanetal2001}.  Whatever
process formed and grew the first massive black holes, it had to be
very efficient.

At the other extreme, we can study the lowest-mass black holes in
galaxy nuclei nearby to us, the left-over seeds that for some
reason never grew to be a billion suns.  As we describe below,
conditions were best to make supermassive black hole seeds soon after
the Big Bang.  Therefore, black holes found in small galaxies today
likely formed early and have not grown significantly since.  If we assume
that black holes form in a similar way in all galaxies, then the
numbers and masses of black holes in small galaxies today contains
clues about the formation of the first black holes
\citep[e.g.,][]{vanwassenhoveetal2010}.  The sheer number of
left-overs will indicate how commonly black hole seeds were formed, as
well as inform future gravitational wave experiments that expect to
see a large number of paired low-mass black holes as they spiral
together and coalesce \citep[][]{hughes2002}.  
Studying the energy output from low-mass black
holes could tell us whether growing black holes at early times
were important in shaping early star formation in the galaxies
around them \citep[e.g.,][]{jeonetal2012}.

Unfortunately, low-mass black holes are difficult to find. Because of 
their low mass, they only have gravitational influence over stars in a 
very small volume at the galaxy center.  Therefore, we are often forced 
to wait until material falls into the black hole.  We detect the black hole 
indirectly via the radiation energy that is released as matter falls in. 

Here we document the last decade of searching for the elusive
low-mass black hole population.  

\section{Formation Paths for the First Massive Black Holes}

To understand the growth of the first supermassive black holes, we
first must determine how the black holes form to begin with.
Theoretically, there are two possible answers.  Either black holes
are created as the end-product of stellar evolution, a process that
continues to produce stellar-mass black holes today, or the black
hole is made directly from the collapse of a gas cloud, which 
requires the high gas fractions and low metallicities of the
early universe.  Once the black hole is formed, it also must
grow. There are likely many growth paths, but a rapid mechanism is
required to explain the $\sim 10^9$~\msun\ black holes that are
observed only hundreds of millions of years after the Big Bang
\citep[e.g.,][]{fanetal2001}.  We first discuss the two formation
routes, and then the possible growth mechanisms. \citet{volonteri2010}
presents a very cogent and recent review of the leading theories for
the formation of the first massive black holes.  I will only
briefly review the subject for completeness, with an emphasis on the
observable consequences at the present day.

\begin{figure*}
\vbox{ 
\vskip -0.truein
\hskip 0.5in
\psfig{file=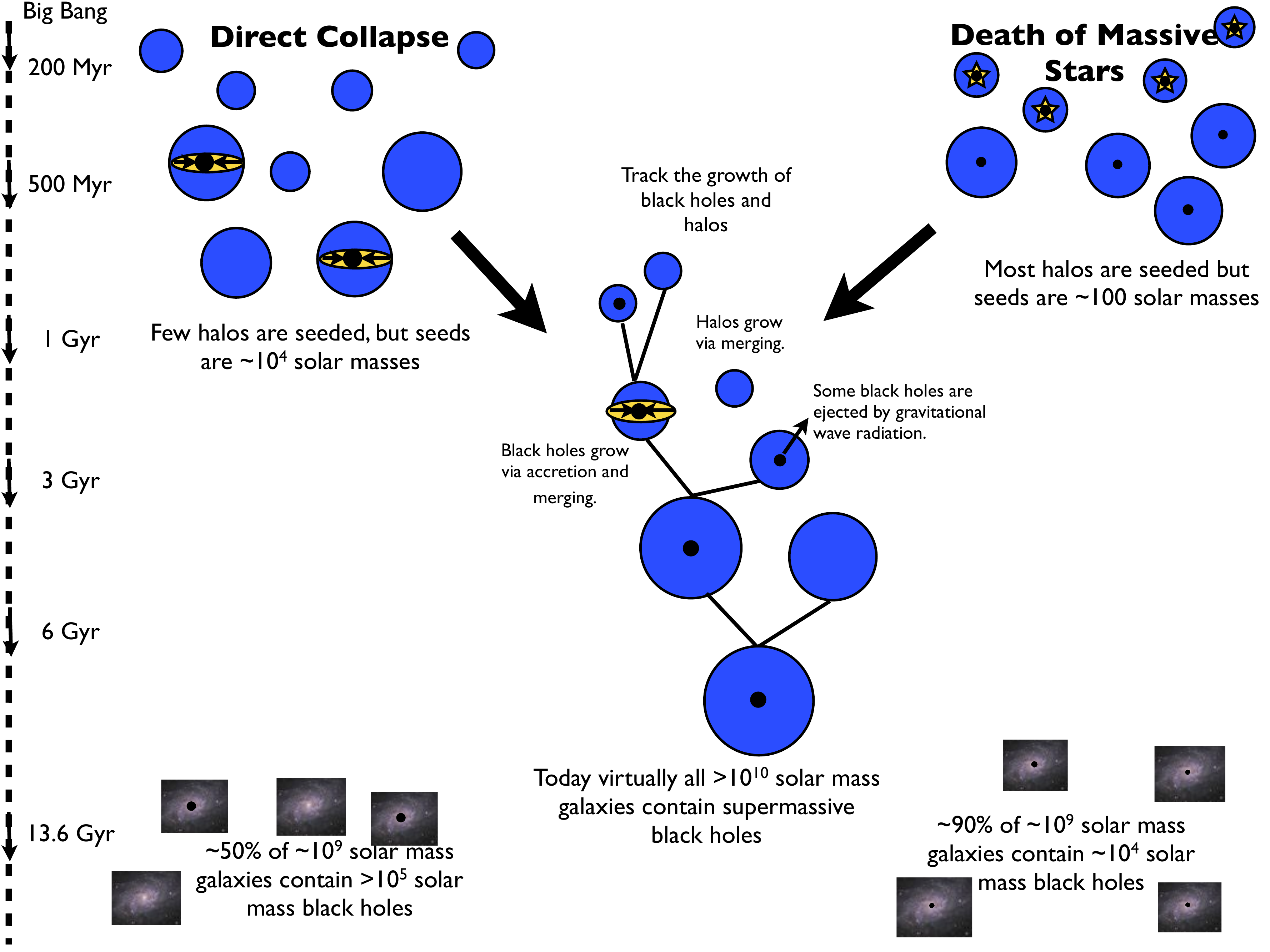,width=0.7\textwidth,keepaspectratio=true,angle=0}
}
\vskip -0mm
\figcaption[]{
Schematic of the evolution of seed black holes assuming two different formation 
mechanisms (the death of the first generation of massive stars vs. the direct collapse 
of gas into a black hole).  Dark matter halos and the galaxies in them grow through 
merging.  Black holes grow both via merging and by accreting gas. 
One additional complication 
is that after merging, gravitational radiation ``recoil'' (see \S 3.1) may send the black 
hole out of the galaxy. At present, we can distinguish between the two scenarios based 
on the fraction of small galaxies that contain massive black holes (we call this the 
``occupation fraction'').
\label{fig:first}}
\end{figure*}

Stellar-mass black holes form when massive stars run out of fuel at
the end of their life.  The first black holes may have formed in the
same way. The first stars were likely very massive
\citep[e.g.,][]{brommyoshida2011}.  In order for stars to form, gas
clouds need to contract; they are able to cool and shrink by emitting
light predominantly in specific element transitions.  Because there
were no elements heavier than He and Li in primordial gas, it was hard
for clouds to cool efficiently. As a result, proto-stars grew much
larger before their gravitational attraction was strong enough to
counteract the internal energy in the gas.  In theory, the end-product
of these massive first stars will depend on the mass.  Stars with
masses less than $\sim 100$~\msun\ or more than $\sim 260$~\msun\ will
make a black hole with a mass approaching that of the star
\citep[e.g.,][]{hegeretal2003}.  For masses in between, it is thought
that pair-instability supernova, in which pair production in the star
center leads to a run-away stellar collapse, will leave no remnant
\citep[e.g.,][]{barkatetal1967}.  Of course, the details of early
stellar evolution are very difficult to test observationally.  There
are many uncertain details, such as whether the first stars formed in
pairs, and how much mass they lose at late stages of evolution.  We
will assume the first stars left behind standard $\sim 100$~\msun\
remnants.

Alternatively, conditions in the early universe may have allowed gas
clouds to collapse {\it directly} into black holes
\citep[e.g.,][]{haehneltrees1993}. Direct collapse requires very low
angular momentum gas that only existed in large quantities soon after
the Big Bang.  In this scenario, only a very small fraction of halos
will manage to form a black hole, and only for a short period of time
soon after the Big Bang.

With these two formation paths in mind, the next question is whether
the black holes created via either path can grow into the very
luminous sources that are observed hundreds of millions of years after
the Big Bang.  With direct collapse models, even in the halos with low
angular momentum content and low molecular hydrogen fraction (and thus
inefficient cooling) the gas will likely settle into a disk, and
require some sort of instability to condense further
\citep{lodatonatarajan2006}. Once sufficiently condensed, the central
$10^4-10^5$~\msun\ of material may very efficiently gain mass as a
dense and round ``quasi-star'', \citep[e.g.,][]{begelmanetal2006}.

It is marginally possible for a stellar-mass seed to grow into a
billion solar mass black hole in hundreds of millions of years, but
only if the black hole manages to grow continuously at the maximal
allowed rate.  Above the so-called ``Eddington'' limit, radiation
pressure forces will exceed gravitational attraction and blow apart
the accretion disk.  In practice, it is difficult for black holes to
grow continuously at their Eddington limit, since the emission from
accretion will heat the gas around the black hole and slow down
subsequent accretion \citep[e.g.,][]{milosetal2009}.  One way to speed
up the growth of black holes created via stellar death is to merge many
smaller seeds into a more massive seed
\citep[e.g.,][]{lietal2007}.  Dense clusters of stars contain many
small seeds that may sink to the center of the cluster and merge to
form a more massive seed with $\mbh \approx 10^4$~\msun\ that can
then grow further into a supermassive black hole
\citep[e.g.,][]{millerdavies2012}.  Similarly, stars may merge first,
forming a supermassive star and then create a more massive seed
\citep[e.g.,][]{portegieszwartetal2004,devecchivolonteri2009}.

\subsection{Observational Consequences}

These different formation scenarios are only interesting if they
predict differences in observations of the real universe.  Eventually,
perhaps with the successor to the \emph{Hubble Space Telescope},
called the \emph{James Webb Space Telescope}, we will detect the
earliest growing black hole seeds \citep[e.g.,][]{brommyoshida2011}.
In the meantime, we can look for clues in how black holes inhabit
galaxies today.  Just looking at supermassive black holes in massive
galaxies provides few insights, because all memory of their humble
beginnings has been erased through the accretion of gas and smaller
black holes.  However, if we focus on the ``left-over'' seeds in small
galaxies (those with stellar masses $M_{\rm gal}<10^{10}$~\msun), the
black holes that never grew, we get a more direct view of
the original seed population \citep[e.g.,][]{vanwassenhoveetal2010}.

Volonteri and collaborators construct models of dark matter halos
merging and growing from the early universe to the present day.  They
put seed black holes into the halos using different prescriptions
depending on how the seeds were formed (see Figure \ref{fig:first}).
Then, they watch the black holes evolve along with the halos.  There
are still many uncertainties associated with these models.  For
instance, as black holes merge, they emit gravitational radiation.  In
general, the gravitational radiation will have a preferential
direction.  When the black holes finally merge, the remnant black hole
will receive a kick from the gravitational radiation that may, in
extreme cases, send the black hole out of the galaxy completely,
called gravitational ``recoil'' \citep[e.g.,][]{merrittetal2004}.
Since it is theoretically uncertain how effective gravitational
radiation will be at ejecting black holes, there is additional
uncertainty added to the models.  Also, the models assume seeds are
formed either via direct collapse or via star death, when in reality
there is likely a mixture.

Given these uncertainties, we focus on the qualitative aspects of the
models.  They predict a higher fraction of low-mass galaxies to
contain nuclear black holes if seeds are created via stellar deaths
(see Volonteri et al.\ 2008 and the purple solid and green dashed
lines, respectively, in Figure \ref{fig:frac}). We are trying to
measure the fraction of galaxies that contain low-mass 
black holes, particularly in host galaxies with $M_{\rm gal} <
10^{10}$~\msun.  As I will show, this work is still in progress.

\section{\it Text Box: Finding Supermassive Black Holes}

\vbox{ 
\vskip -0.truein
\hskip 0.in
\psfig{file=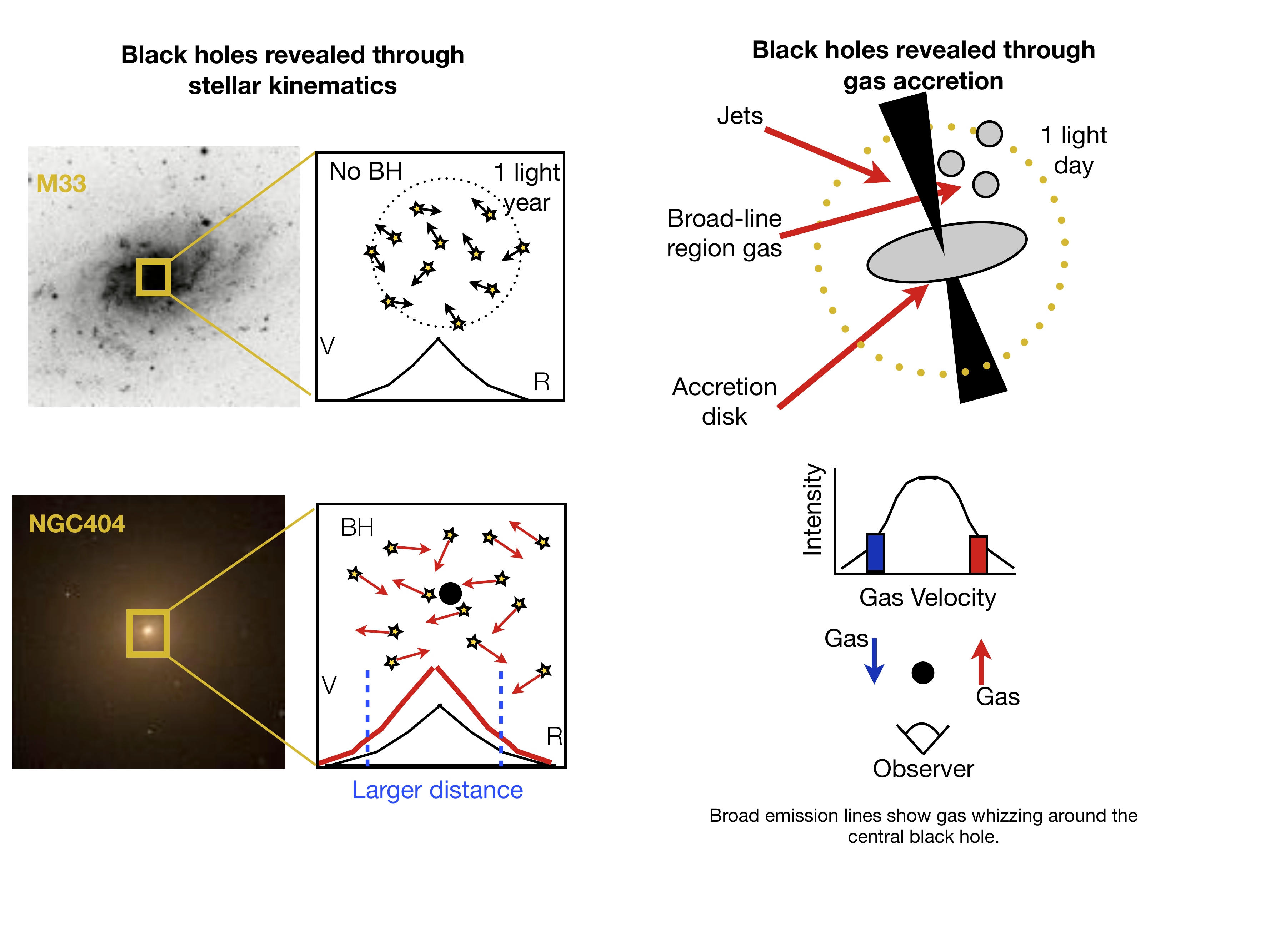,width=0.45\textwidth,keepaspectratio=true,angle=0}
\label{fig:detection_methods}}

\subsection{Direct Detection with Stellar Dynamics}

The most direct way to demonstrate that a supermassive black hole
exists at the center of a galaxy is to look for the evidence of the
gravity of the black hole from the motions of gas or stars at the
galactic nucleus. Just like planets going around the sun, we can use
the laws of gravity to translate the average velocities of stars
around the black hole into a mass.  At the center of our own 
Milky Way galaxy, researchers have charted the motions of individual stars
whipping around at the galaxy center for over a decade.  The star
motions provide unambiguous evidence for a $4 \times 10^6$ solar mass
black hole \citep{ghezetal2008,gillessenetal2009}.

In other galaxies, we cannot study individual stars.  We can, however,
still see the signature of the black hole in the average star motions
near the galaxy center.  Stars move much faster on average if there is
a black hole at the center of the galaxy then if there is not. The
bigger the black hole, the faster the average motions of the
stars. Because the black hole comprises only a fraction of a percent
of the total mass of the galaxy, only the stars or gas very near the
galaxy center can feel the gravitational attraction of the black hole.
To detect these fast-moving stars requires either the high spatial
resolution of the \emph{Hubble Space Telescope} or adaptive optics
from the ground.  In addition to stars, orbiting gas clouds can also
be used to weigh the black hole, using very similar principles 
\citep[e.g.,][]{barthetal2001,herrnsteinetal2005,kuoetal2011}.  

It is only possible to detect the gravity of big black holes that are
in relatively nearby galaxies.  As shown in the bottom panel of Figure
\ref{fig:detection_methods}, the signal from the gravity of the black
hole is strongly concentrated towards the galactic center.  When the
galaxy is further away, the motions of stars near the black hole are
blurred together with more distant stars that don't feel the black
holes gravity, making its influence imperceptible.  Similarly, as the
black hole mass gets smaller, the high velocity stars become more and
more difficult to detect.  To study low-mass black holes, we are
generally forced to wait until matter falls into the black hole and forms an
accretion disk, which we can detect.

\subsection{Detecting accretion onto black holes}

Occasionally, gas will make its way into the galaxy nucleus and into the
black hole.  However, the gas must dissipate its energy and
angular momentum to fall into the deep gravitational
potential of the black hole. Nature uses accretion disks to funnel matter into 
black holes; accretion disks are many hundreds to
thousands of times smaller than the gas disks described
in \S4.1. Accretion disks are also hot, and radiate most of
their energy in the ultraviolet.  While we cannot see the peak
of the radiation from the accretion disk, we can recognize the
following signatures of an accreting black hole:

\noindent
$\bullet$ Unresolved X-ray emission at the galaxy nucleus is a sign of
an accreting black hole.  High-energy interactions between photons and
electrons form an X-ray corona above the accretion disk.  Because the
corona region is very compact, the light signals propagate from one
side of the corona to the other rapidly, enabling variability on short
timescales.  Other processes can create X-rays in galaxies, but the
X-rays are generally of lower energy, relatively fainter, and do not
tend to vary on the rapid timescales seen in accreting black holes.

\noindent
$\bullet$ Sometimes accretion onto central black holes is accompanied by
  jets of accelerated particles that emit radio waves.  Jet emission
  often accompanies accretion disks, although the exact mechanisms
  responsible for launching the jets in accreting supermassive black
  holes are not fully understood.

\noindent
$\bullet$ Outside the accretion disk is gas that orbits the black hole and
  emits spectral line transitions: for instance hydrogen
  atoms emit optical light when electrons fall into the second energy
  level.  Intrinsically, light from electron transitions are emitted
  at a specific frequency.  However, because the gas is moving
  towards and away from us, the frequency we observe is shifted to the
  blue or the red via the Doppler shift.  The faster the gas is
  moving, the wider is the range of velocities that we observe in line
  emission.  The fast-moving gas that orbits close to the black hole but
  outside the accretion disk is called the ``broad-line'' region
  because of the high observed velocities.

\noindent
$\bullet$ The gas in the galaxy on larger scales is also
  illuminated by emission from the accretion disk.  Because the
  accretion disk emits strongly in the ultraviolet and X-ray, the gas
  in the galaxy is excited to a wide range of temperatures.  Gas that
  is excited by accretion shows specific fingerprints in the ratios of
  different atomic transitions.  These emission line 
  fingerprints can be observed in the ultraviolet, optical, near-infrared 
  and mid-infrared wavelengths \citep[e.g.,][]{hfs1997spec}. For instance,
  emission lines from quadrupley ionized Ne are typically only excited
  in the vicinity of an accreting black hole \citep{satyapaletal2007}.

\subsection{Using accretion to determine black hole mass}

In cases for which we cannot determine a black hole mass directly using the
motions of stars or gas, we can get an approximate idea of the mass by
observing the motions of gas clouds in the broad-line region.  We use
the gas clouds in a very similar way to the stars: the faster the gas
moves on average, the wider the range of velocities we observe in the gas.
The wider the range of observed velocities, the more massive the black
hole. However, we need to know not only how fast the gas clouds are
moving, but also how far away they are from the central black hole.
Thinking about the planets in the solar system as an analogy, we
know that Pluto moves much more slowly around the Sun than the Earth
just because it is at a much larger distance from the Sun.  In the
case of the gas clouds, it is actually quite challenging to determine
their distances from the black hole.  In some cases the time
delay in variable emission from the accretion disk itself, and from
the broad-line region further away, provides a size scale for the
emission region, because the distance is just the delay time times the
speed of light.  Using this technique, the mass of the black hole in
NGC 4395 is found to be $\mbh~\approx 10^5$~\msun\
\citep{petersonetal2005,edrietal2012}.  Measuring these time delays is
very time consuming.  Usually, we do not measure the broad-line
region size directly.  Instead, we use a correlation between the
luminosity of the black hole and the broad-line region size to
estimate a radius for the broad-line region gas.

\section{The search for low-mass black holes}

Astrophysical supermassive black holes were first discovered as
``QSOs'' -- quasi-stellar objects with very high intrinsic
luminosities and very small sizes \citep[e.g.,][]{schmidt1963}.  By
the late 1970s there was compelling evidence that QSOs are powered by
accretion onto a supermassive black hole.  They were called ``active
galactic nuclei'' (AGN) because they were shining via the energy
released as material falls into (or is accreted by) the central
supermassive black hole \citep{lyndenbell1969}.  The existence of real
supermassive black holes, with masses of hundreds of millions of times
the mass of the sun became commonly accepted, but it was far less
clear whether these ``monsters'' represented a rare and long-lived
phenomenon, or whether all galaxies contained supermassive black holes
with short-lived bright episodes.

\begin{figure*}
\vbox{ 
\vskip -0.truein
\hskip 20mm
\psfig{file=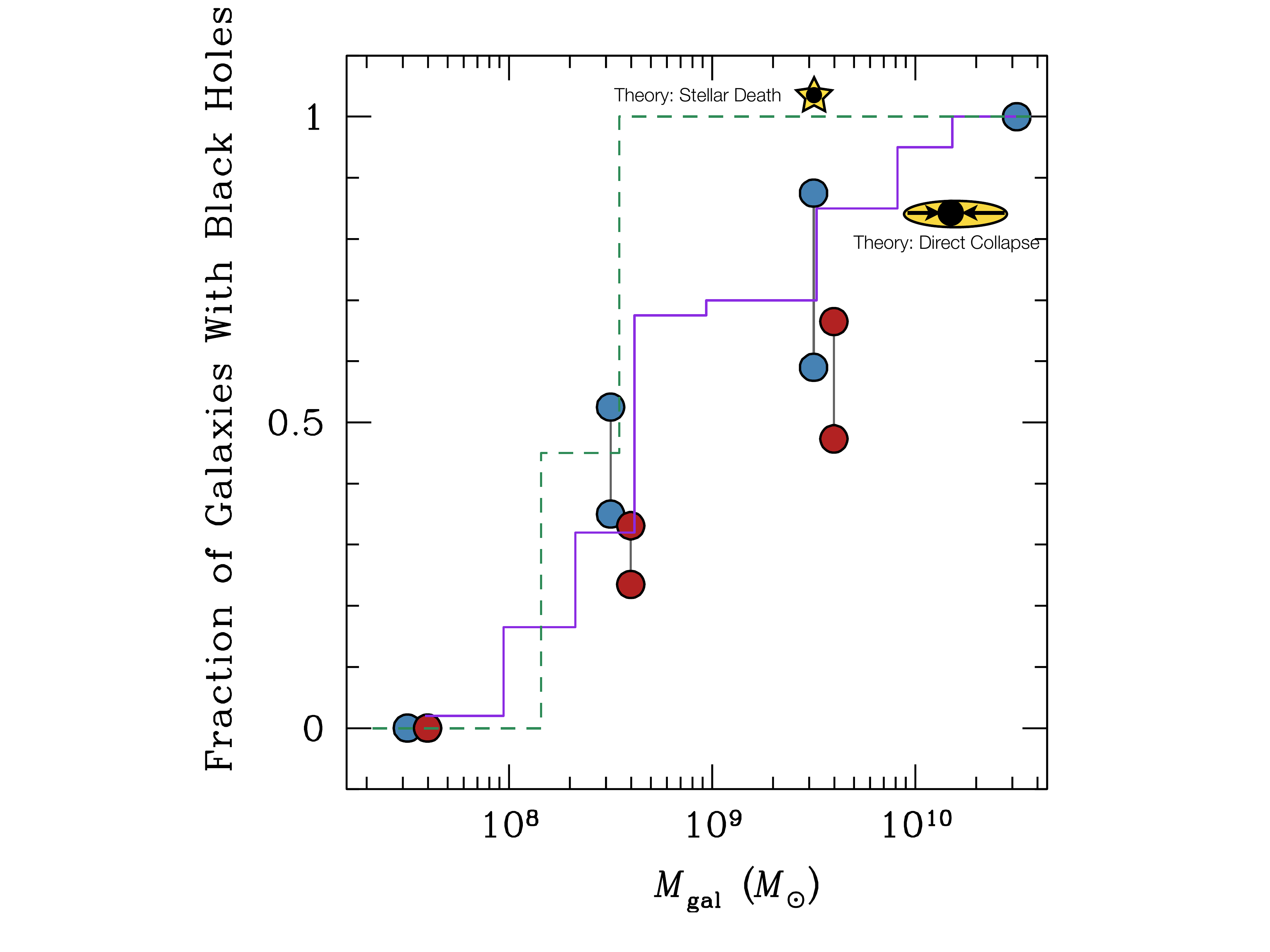,width=0.8\textwidth,keepaspectratio=true,angle=0}
}
\vskip -0mm
\figcaption[]{
  We show the expected fraction of galaxies with $M_{\rm gal} \lesssim
  10^{10}$~\msun\ that contain black holes with $\mbh \gtrsim 3 \times
  10^5$~\msun, based on the models of \citet{volonterietal2008}, as
  presented in \citet{volonteri2010}, for high efficiency massive seed
  formation (solid purple line) , as well as stellar deaths (greed
  dashed line).  From data in the literature, in large circles we show
  the fraction of galaxies containing black holes greater than
  $10^6$~\msun\ (lower points) and greater than $3 \times 10^5$~\msun\
  (higher points) based on the paper by \citet{desrochesho2009} (blue)
  and \citet{galloetal2010} (red).  See text for details.  Although
  the uncertainties are very large, we find tentative evidence in
  support of the efficient massive seed models (purple solid line).
\label{fig:frac}}
\end{figure*}

Twenty years later we finally learned that supermassive black holes
are common.  In fact, we believe that most massive galaxies contain a
central supermassive black hole.  The evidence came from both stellar
dynamics and accretion (see text box).  A survey by Ho, Filippenko, \&
Sargent \citep{hfs1997spec} searched the centers of nearby, ``normal''
galaxies for subtle evidence that trace amounts of gas was falling
into a central black hole.  Amazingly enough, most ($\sim 70\%$)
massive galaxies showed clear signs of accretion onto a supermassive
black hole \citep[see review in][]{ho2008}.  At the same time
stellar dynamical work was providing increasing evidence that every
bulge-dominated galaxy harbors a supermassive black hole
\citep[e.g.,][]{richstoneetal1998}.  It became clear that black holes
were preferentially associated with galaxy bulges\footnote{Bulges
  are ellipsoidal in shape and comprised of mostly old stars that move
  on random orbits through the galaxy.  In contrast, disks are flat
  components of galaxies, where stars all orbit the galaxy on coplanar
  circular paths.  Disks contain gas and ongoing star formation.  If a
  bulge component contains no disk, we call it an elliptical galaxy.}.
Furthermore, the ratio of black hole to bulge mass was apparently
constant to within a factor of two to three
\citep[][]{tremaineetal2002}.

Unfortunately, understanding the black hole population becomes
increasingly challenging as one considers lower and lower-mass
galaxies.  Low-mass galaxies typically contain more cold gas, more
dust, and higher levels of ongoing star formation.  The dust obscures
emission from accretion, while the star formation masks it.
Furthermore, if the correlation between BH mass and bulge mass
applies, the BHs in smaller galaxies are less massive, which makes their 
emission weaker.  Lower-mass black holes also exert a
smaller gravitational force, so that it becomes more and more
challenging to detect stars moving under the influence of the black
hole.

\begin{figure*}
\vbox{ 
\vskip -0.truein
\hskip 0.5in
\psfig{file=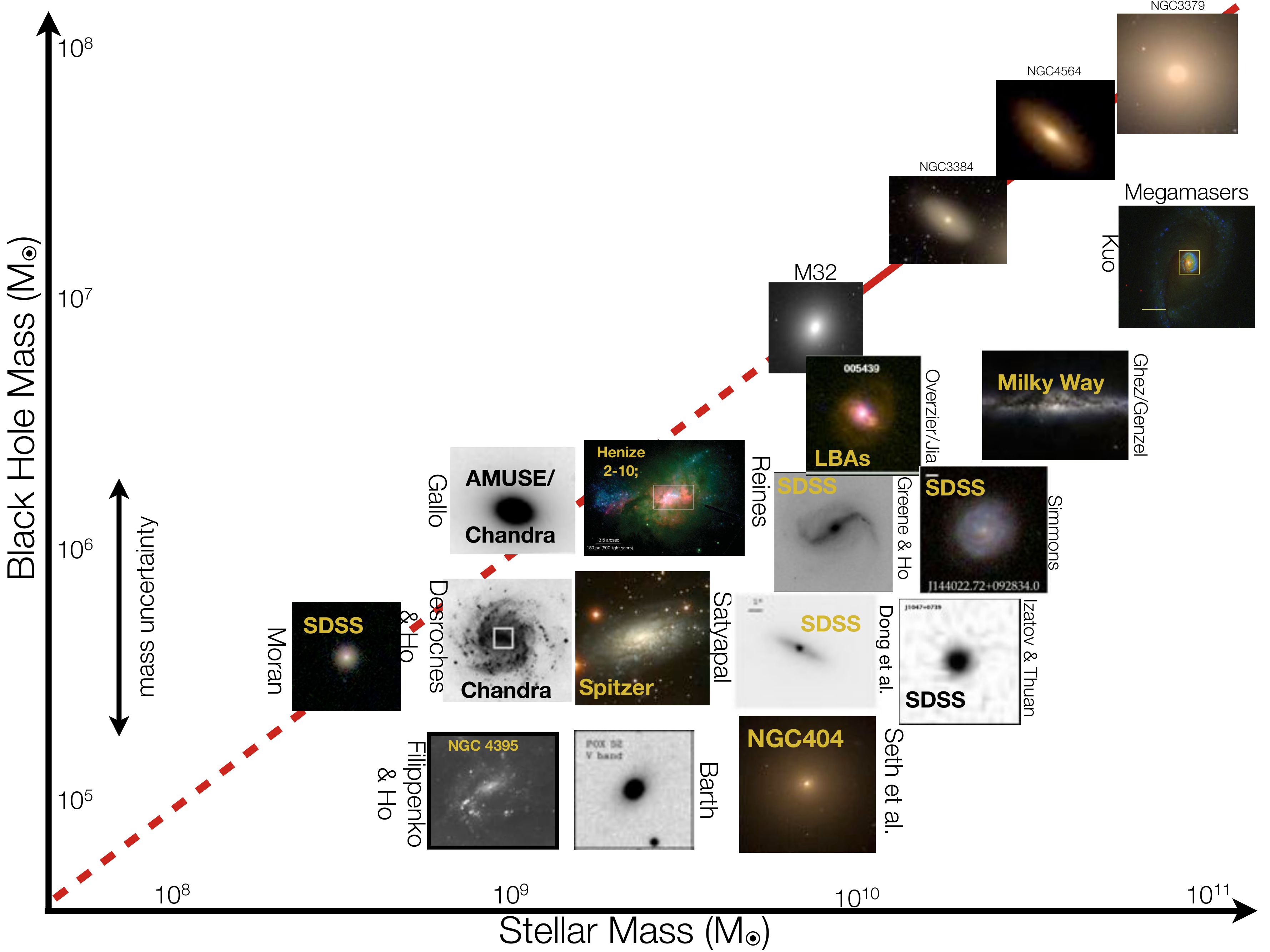,width=0.7\textwidth,keepaspectratio=true,angle=0}
}
\vskip -0mm
\figcaption[]{
The relationship between galaxy bulge mass and black hole mass is linear for 
bulge-dominated galaxies, as shown by the solid red line 
in the upper right from \citet{haeringrix2004}.  To guide the eye, we have extrapolated 
this relationship down to lower black hole masses with the dotted red line.
However, in disk-dominated galaxies, particularly at low mass, there 
is no tight correlation between \mbh\ and properties of the galaxy.  We illustrate the 
wide range of galaxy types hosting low-mass black holes, roughly placed 
in accordance with galaxy and black hole mass.  However, note that only NGC 404 
has a direct dynamical black hole mass measurement.
In all other cases, the black hole masses are very approximate, as illustrated by 
the error bar to the left.
\label{fig:galbh}}
\end{figure*}

\subsection{Dynamical Evidence for Black Holes in Low-mass Galaxies}

For the handful of bulgeless galaxies nearest to us, it is possible to
search for the gravitational signature of a central black hole. In
stark contrast to bulge-dominated galaxies, these nearby bulgeless
galaxies show no evidence for a central massive black hole, with an
upper limit of $1500$~\msun\ for the galaxy M33
\citep{gebhardtetal2001} and of $10^4$~\msun\ for NGC 205
\citep{vallurietal2005}.  \citet{loraetal2009} and
\citet{jardelgebhardt2012} find that the black hole masses of two very
low-mass dwarf galaxies in our local neighborhood cannot be larger
than $\lesssim10^4$~\msun.  While massive, bulge-dominated galaxies
contain black holes (an ``occupation fraction'' close to unity),
clearly the occupation fraction drops in galaxies without bulges. But,
do $50\%$ of dwarf galaxies contain black holes or only $1\%$?  And
how does that fraction change with the mass of the galaxy?

Apart from M33 and NGC 205, there are very few galaxies near enough to
place interesting limits on the presence or absence of a black hole based on
the motions of stars or gas at the galaxy center (see text box above).
\citet{barthetal2009} studied the nuclear kinematics
of the bulgeless galaxy NGC 3621.  This galaxy also shows some evidence for
accretion (see \S 3.3).  They place a very conservative upper limit of
$3 \times 10^6$~\msun\ on the mass of a central massive black hole, which
would be improved by better measurements of the stellar ages in the
star cluster surrounding the black hole. \citet{sethetal2010} used the motions
of a gas disk in the center of the small S0 galaxy NGC 404 to find a
likely black hole mass of $\sim 5 \times 10^5$~\msun.
\citet{neumayerwalcher2012} find upper limits of $\sim 10^6$~\msun\ for
nine bulgeless spirals, confirming that such galaxies contain low-mass
black holes if they contain a central black hole at all. 

\subsection{Bulgeless Galaxies with Active Nuclei}

M33 taught us that not all low-mass galaxies contain central
supermassive black holes.  The galaxy NGC 4395, a galaxy very similar
in mass and shape to M33, shows that some galaxies without bulges do
contain nuclear black holes \citep{filippenkosargent1989}.  Like M33,
NGC 4395 is small and bulgeless.  Unlike M33, NGC 4395 contains
unambiguous evidence for a central massive black hole (see text box),
including extremely rapid variability in the X-rays
\citep{shihetal2003} and a radio jet \citep{wrobelho2006}.  While we
do not know precisely, the black hole mass is likely
$10^4-10^5$~\msun\
\citep{filippenkoho2003,petersonetal2005,edrietal2012}.

NGC 4395 highlights the utility of using nuclear activity as a
fingerprint of low-mass black holes when their gravitational signature
is undetectable.  In 2004, Aaron Barth reobserved the forgotten active galaxy
POX 52 \citep{kunthetal1987,barthetal2004}, which has a near-identical optical
spectrum to NGC 4395.  POX 52 also appears to contain a $\sim
10^5$~\msun\ black hole.  NGC 4395 went from being an unexplained
oddball to the first example of a class of objects, with POX 52 being
the second example. But are there more? We were inspired to perform
the first large systematic search for this new class of ``low-mass''
accreting black holes.  

In 2003, the Sloan Digital Sky Survey \citep[ SDSS][]{yorketal2000}
had just started to provide pictures and spectra of objects over
one-quarter of the sky, exactly what we needed to search for the rare
and elusive low-mass black holes.  We decided to find accreting
black holes and use the motions of gas very close to the black hole to
trace the black hole mass (see \S 4.2).  We went through hundreds of
thousands of galaxy spectra to pick out the accreting black holes with
fast-moving gas (\S 4.2).  We then picked out the $\sim 200$ systems 
with masses $<10^6$~\msun\ \citep{greeneho2004,greeneho2007}.
We chose this mass because it is similar to the mass of the black hole
at our Galaxy center, and serves as the anecdotal low-mass cut-off 
of supermassive black holes. \citet{dongetal2012} also searched through the
SDSS for low-mass black holes with similar criteria, increasing the
total sample by $\sim 30\%$.

Subsequent searches of the SDSS have adopted different approaches.
While looking for galaxies with pristine, metal-free gas,
\citet{izotovthuan2008} present evidence for accreting black holes in
four vigorously star forming galaxies, again based on the
detection of fast-moving gas that is most likely orbiting a black hole
with $\mbh \sim 10^4-10^6$~\msun\ (see
Fig. \ref{fig:detection_methods}).  In contrast, a number of groups
are now first selecting low-mass galaxies, and then searching for
signatures of accretion \citep[e.g.,][]{barthetal2009}.

\subsection{Multiwavelength Searches}

Searches for low-mass black holes using the SDSS were an important
first step, and allowed us to comb through hundreds of thousands of
galaxies.  However, they are fundamentally
limited in two ways.  Firstly, in galaxy nuclei with ongoing star
formation, dust obscuration and emission from star formation hides the
evidence of nuclear activity.  Secondly, the SDSS takes spectra of a
biased sample of relatively bright galaxies, which makes it very
difficult to calculate a meaningful occupation fraction
\citep{greeneho2007b}.

An obvious way to circumvent these biases, and complement the original
optical searches, is to use other wavebands.  X-rays, for instance
(see text box), are such high-energy photons that they can only be
hidden by very large quantities of gas.  Radio and mid-infrared
wavelengths are also relatively unaffected by dust absorption.  On the
other hand, multiwavelength searches to date have been restricted to
small samples.

The fraction of low-mass ($M_{\rm gal} < 10^{10}$~\msun) galaxies with
X-ray emission coming from the nucleus has been studied both as a
function of stellar mass \citep{galloetal2010,milleretal2012} and
galaxy morphology \citep{desrochesho2009,ghoshetal2008}.  The former
studies focused on galaxies comprised of old stars, while
the later focused on star-forming galaxies.  Less than $20\%$ of 
non-starforming galaxies with $M_{\rm gal} < 10^{10}$~\msun\ have
nuclear X-ray sources with $L_{\rm X} \gtrsim 2.5 \times
10^{38}$~erg~s$^{-1}$, while in star forming galaxies of similar mass,
$\sim 25\%$ of galaxies contain X-ray nuclei above the same
luminosity. The difference in detection rate is likely due to a lack
of gas to consume in the red galaxies. In \S 6, I will use these
detection fractions to estimate the occupation fraction in galaxies
with $M_{\rm gal} < 10^{10}$~\msun.

The X-ray luminosities probed here are very low.  If a $10^5$~\msun\
black hole has very little to accrete, it will only shine very weakly,
in this case with an X-ray luminosity as low as $L_X \lesssim 10^{38}$
erg~s$^{-1}$.  However, stellar mass black holes can sometimes shine
with this luminosity as well, although they are not very common; the
best guess is that $\sim 10\%$ of the sources are actually powered by
$10$~\msun\ black holes while the rest are powered $\sim 10^5$~\msun\
black holes \citep[e.g.,][]{galloetal2010}. However, as we look at
less and less luminous X-ray sources, more and more of them will be
powered by stellar-mass black holes, until for $L_X \lesssim 10^{37}$
erg~s$^{-1}$, nearly all the sources will be stellar-mass black holes.
Due to confusion about the nature of the detected objects, we are
reaching the limit of what X-ray searches alone can tell us about the
demographics of low-mass black holes.  It is possible that including
X-ray variability information will also help weed out the stellar-mass
black holes \citep{kamizasaetal2012}.

The high-ionization [Ne {\small V}] line, detected in mid-infrared
spectroscopy, is a reliable indicator of AGN activity since starlight
likely cannot excite this transition
\citep[e.g.,][]{satyapaletal2007}.
\citet{satyapaletal2008,satyapaletal2009} focus on galaxies with
little to no bulge component.  In galaxies with very small bulges,
they find a detection fraction similar to the X-ray studies
\citep[$\sim 20\%$; see also][]{gouldingetal2010} but in the galaxies
with no bulge whatsoever, their detection fraction drops nearly to
zero (one galaxy out of 18 contains a [Ne {\small V}] detection).
While the precipitous decline of detected bulgeless galaxies provides
evidence for a dramatic decline in the occupation fraction of
bulgeless galaxies, it is worth noting that the observations of the
latter galaxies were not as sensitive.

Further progress requires observations at multiple wavelengths. For
example, \citet{reinesetal2011} identified a likely $10^5-10^6$~\msun\
black hole in the center of the low-mass star-forming galaxy Henize
2-10.  Radio or X-ray emission alone would have been unconvincing,
since either could easily be explained by processes relating to star
formation.  However, the spatial coincidence of the radio and X-ray
source, their relative brightness, and their distance from clusters of
forming stars make a compelling case for a low-mass black hole in this
galaxy.  A larger sample of galaxies like Henize 2-10
\citep[][]{overzieretal2009} also appear to have accreting black holes in
some cases.  Again, it is the combination of X-ray and radio
detections that argues for black holes in these galaxies
\citep{jiaetal2011,alexandroffetal2012}.

\section{Properties of known low-mass black holes and their host galaxies}

In addition to measuring the fraction of low-mass galaxies that
contain black holes, it is of interest to determine whether black
holes with lower mass emit a different spectrum than more massive
accreting black holes.  For instance, one naively expects that the
accretion disk will get physically smaller and thus hotter as the
black hole mass decreases.  In turn, gas in the vicinity of the
black hole will be heated by more energetic photons.  Predictions of
the impact of low-mass black holes on the gas conditions in the early
Universe require empirical measurements of the radiation from low-mass
black holes.

In practice, since accretion disk emission around supermassive black
holes peaks in the ultraviolet, it is difficult to unambiguously
measure changes in the disk temperature with mass
\citep[e.g.,][]{davisetal2007}.  We have measured a few intriguing
properties of the radiation from low-mass black holes, although thus far 
we have studied only the most luminous of them.  First, they
appear to have a very low incidence of jet activity
\citep{greeneetal2006}.  Second, we see indirect evidence that they
have hotter accretion disks than their more massive cousins
\citep{greeneho2007chandra,desrochesetal2009,ludwigetal2012,dongetal2012}
as expected from basic disk models \citep{doneetal2012}.  As the
accretion disk gets hotter, its impact on the surrounding gas will
grow.  Thus, growing black holes may well impact the formation of the
first stars and galaxies \citep[e.g.,][]{jeonetal2012}.

Accretion onto a central black hole has been found in low-mass
galaxies of all shapes and with all levels of ongoing star formation.
\citet{milleretal2012} have found accreting black holes in galaxies
comprised predominantly of old stars \citep[see
also][]{pellegrini2010}, while \citet{izotovthuan2008},
\citet{reinesetal2011}, and \citet{jiaetal2011} have reported evidence
of accreting black holes in vigorously star forming galaxies.  Some
host galaxies are round \citep{barthetal2004}, while others are pure
disks \citep{filippenkoho2003}.  We display the variety of host galaxy
morphologies in Figure \ref{fig:galbh}.

Supermassive black holes in bulge-dominated galaxies obey remarkably
tight correlations between black hole mass and the properties of the
host galaxy.  For a long time it was simply unknown whether the
correlations seen for bulges apply to disk-dominated galaxies;
dynamical black hole mass measurements in disk galaxies are severely
compromised by the presence of dust and young stellar populations.
Early on, based on very indirect arguments, we saw evidence that the
relationship between black holes and galaxies extended to low-mass and
even bulgeless systems \citep{barthetal2005}.  However, as the number
of available dynamical black hole masses in disk galaxies grows, it
becomes increasingly clear that disk-dominated galaxies do not obey
tight scaling relations with the central supermassive black hole
\citep{hu2008,greeneetal2010,kormendyetal2011}.  Apparently the
physical process that builds galaxy bulges (the merging of galaxies,
we think), is also instrumental in growing black holes and
establishing the scaling relations between black holes and bulges
\citep[e.g.,][]{mihoshernquist1996}.

\section{Occupation Fractions}

Let us now determine whether existing observations of low-mass black
holes favor a particular formation route for primordial seed black
holes.  I want to estimate the fraction of galaxies containing black
holes as a function of galaxy mass: the occupation fraction.  Based on
previous work, we assume that all galaxies with stellar mass $M_{\rm
  gal} > 10^{10}$~\msun\ contain black holes.  To study the occupation
fraction in lower-mass galaxies, I will use the two X-ray studies
discussed above from \citet{desrochesho2009} and \citet{galloetal2010}
combined with \citet{milleretal2012}.  Note that existing optical
studies, while they include a larger number of objects, cannot be used
for measuring occupation fractions because of their bias towards
luminous host galaxies. While the samples with X-ray measurements are
smaller, they should be unbiased.

Since we are using accretion to discover black holes (via X-ray
emission), we will not detect all black holes.  For instance, we may
detect X-rays from $10\%$ of galaxies with $M_{\rm
  gal}=10^{11}$~\msun, while they all contain supermassive black
holes.  If we then detect X-rays from $1\%$ of galaxies with $M_{\rm
  gal}=10^{8}$~\msun, we can conclude that only $10\%$ of $M_{\rm
  gal}=10^{8}$~\msun\ contain black holes\footnote{We are making an
  assumption that the fraction of active black holes is always the
  same (in my example $10\%$), independent of the galaxy or the black
  hole mass.  This is a strong assumption that we must make in order
  to proceed, but in detail it is probably incorrect. For instance,
  galaxies with less gas may be less likely to harbor an active black
  hole}.  For each X-ray sample we need a comparison sample of more
massive galaxies with similar X-ray observations.  The Gallo et al.\
sample spans a wide range in stellar mass, so the comparison sample of
bulges is built in.  As a complementary sample of more massive disk
galaxies to compare with the Desroches sample, I take the archival
X-ray survey of \citet{zhangetal2009}. Given the existing X-ray
surveys, we are not sensitive to black holes with masses $\mbh
\lesssim 3 \times 10^5$~\msun.

The only additional complication is that lower-mass black holes are
intrinsically more difficult to detect; their maximum luminosity gets
lower as the black hole mass goes down.  In other words, the Eddington
limit (where the radiation pushing out balances the pull of gravity)
increases linearly with black hole mass.  At the luminosity limit of
my survey, I can see massive black holes to a much lower fraction of
their maximum luminosity than low-mass black holes.  To remove this
bias, I will normalize all the observed X-ray luminosities to the
Eddington limit of that black hole ($L_X/L_{\rm Edd}$), and then only
consider detections down to a fixed $L_X/L_{\rm Edd}$ limit.  In most
cases I do not have direct measurements of black hole mass.  Galaxies
with $M_{\rm gal} > 10^{10}$~\msun\ are assigned a black hole mass
based on the correlation between galaxy mass and black hole mass,
while I assume no such correlation holds for galaxies with $M_{\rm
  gal} < 10^{10}$~\msun. Instead, I just assume that every black hole
has a mass between $3 \times 10^5$ and $10^6$~\msun.  The limiting
$L_X/L_{\rm Edd}$ is calculated accordingly in each
case\footnotetext{To calculate uniform stellar mass for all galaxies
  in both samples, I follow \citet{galloetal2010} and use correlations
  between galaxy $B-V$ color and the mass-to-light ratio of the stars from
  \citet{belletal2003}.}.

The results are shown in Figure \ref{fig:frac}.  The Desroches \& Ho
limits are shown in blue, while the Gallo points are shown in red, for
$10^6$~\msun\ (lower) and $3 \times 10^5$~\msun\ black holes (upper).
We have no constraints on black holes of lower mass yet.  The
expectations from theory for seeds created via stellar death (green
dashed) and direct collapse (solid purple) are shown as
well. Obviously, the limits are not yet good enough to say anything
definitive, but tentatively the data seem to prefer massive seed models.

Improving these limits will require a multi-pronged approach.  As
argued above, we are reaching the limit of utility of X-ray surveys
because we run into confusion from stellar-mass black holes.  However,
X-ray surveys that stare at the same part of the sky for a very long time 
are starting to uncover accreting black holes in 
galaxies with $M_{\rm gal} < 10^9$~\msun. At the same time, combining 
sensitive radio and X-ray surveys may yield interesting new constraints on 
black holes at much lower fractions of their Eddington luminosity.

\section{\it Text Box: Thinking outside the nucleus: 
Off-nuclear black holes and stellar clusters}

We focus here on black holes found at galaxy centers.  However, we
want to highlight two other very interesting locations that may harbor
hitherto unknown low-mass black hole populations: the centers of dense
stellar clusters, and in the outer parts of galaxies.

Another route to seed black hole growth may occur if the
original seed is formed in the center of a dense cluster of stars.
The black hole might grow to $10^4-10^6$~\msun\ by accreting
smaller black holes at the centers of dense stellar clusters
\citep[e.g.,][]{ebisuzakietal2001}. Then we may expect to find $\sim
10^4$~\msun\ black holes at the centers of globular clusters.  These
dense clusters of stars comprise some of the oldest stellar systems in
the universe.  Despite many searches for black holes in stellar
clusters, both using dynamical techniques
\citep[e.g.,][]{vandermareletal2002,gerssenetal2003} and looking for
signatures of accretion \citep[e.g.,][]{maccaroneetal2005}, there is
not yet definitive evidence for black holes in globular clusters.

There is one exception.  The most massive globular clusters, we
believe, were formed not as isolated clusters of stars but rather as
the nucleus of a galaxy that was then torn to shreds as it was eaten
by a larger galaxy.  These most massive clusters also tend to show a
wider range of stellar age and chemical compositions than is seen in
typical globular clusters, suggesting that they were formed gradually
rather than as a single unit.  The three prominent stripped galaxy
candidates that are in our own local neighborhood all show dynamical
evidence for a central $10^4$~\msun\ black hole
\citep[][]{gebhardtetal2005,noyolaetal2010,ibataetal2009}, although
the detections are still controversial
\citep{vandermarelanderson2010}.  Radiation from these putative black
holes has not yet been detected
\citep{wrobeletal2011,millerjonesetal2012}.

Another suggestive line of evidence for black holes in the centers of
star clusters come from the intriguing ``Ultra-luminous'' X-ray
sources.  As the name implies, these targets have very high X-ray
luminosities, so high that they exceed the maximum (Eddington)
luminosity for a stellar-mass black hole.  An easy way to explain the
high luminosities is to power these X-ray sources with
intermediate-mass black holes with masses of $100-10,000$~\msun.  The
evidence is not ironclad, however, since it is possible to reproduce
the properties of ULXs with stellar-mass black holes in all but a few
extreme cases \citep[e.g.,][]{socratesdavis2006}.  Unfortunately,
determining the masses of black holes that power ULXs has proven
impossible to date.

There is a spectacular ULX that deserves mention.  ESO 243-49 HLX-1
has an X-ray luminosity greater than $10^{42}$, the Eddington
luminosity for a $10^5$~\msun\ black hole \citep{farrelletal2009}.
The X-ray source is found offset from the main body of the galaxy ESO
243-49, but at the same distance as the galaxy \citep{wiersemaetal2010}.  HLX-1 is
likely embedded in a stellar cluster with $M_{\rm gal} \approx 10^6$~\msun\
\citep{farrelletal2012}, perhaps the remnant of a galaxy that was
eaten by ESO 243-49 in the past.  ESO 243-49 HLX-1 is an
intriguing source, but so far no other sources like it are known. 

It is quite possible that many intermediate-mass black holes may
reside outside of galaxy nuclei.  As galaxies merge, they acquire
black holes as well as stars.  Many of these black holes may never 
reach the galaxy center, but reside in
galaxy halos \citep[e.g.,][]{islametal2003}, where they would be very 
difficult to find.  If the
black hole is massive enough to sink to the galaxy center, then it may
merge with an existing black hole; the resulting gravitational
radiation could in principle eject the black hole from the galaxy
\citep[e.g.,][]{merrittetal2004}.  At higher redshift, the number of
infalling accreting black holes is high \citep{comerfordetal2009},
while at low redshift, systems like ESO 243-49 HLX-1 appear to be
rare.

\section{Future Prospects}

How can we make progress on determining the space density of the
lowest-mass black holes?  At the moment, we are limited by the largest
distances that we can probe in unbiased samples.  Optical
spectroscopic surveys such as the SDSS have yielded large samples, but
with selection biases that are difficult to quantify.  Searches in
other wavebands, while cleaner in terms of selection effects, reach
limited distances and thus contain small numbers of objects.  I see
multiple paths forward.  The first is to look harder for black holes
in local galaxies.  We are reaching a fundamental limit in using
X-rays, since stellar-mass black holes will dominate the emission in
surveys that push an order of magnitude deeper \citep{galloetal2010}.
On the other hand, the increased sensitivity of radio telescopes
(particularly the Jansky Very Large Array; Jansky VLA) open the possibility of a combined radio
and X-ray survey.  While on an object-by-object basis there still may
be complications \citep[e.g.,][]{millerjonesetal2012}, the combination
of a radio and X-ray source will be compelling evidence for a low-mass
black hole.  Likewise, we may be able to rely on variability (both in
the X-ray and in the optical) in the future \citep{kamizasaetal2012}.

The second path is to try to find more accreting black holes in small
galaxies by searching over larger distances rather than by looking for
fainter sources.  Very sensitive X-ray surveys \citep{xueetal2011}
should allow such experiments, while the newly refurbished Jansky VLA
could perform a very sensitive search using radio wavelengths.  We
also need better measurements of what fraction of more massive
galaxies contain accreting black holes, as our comparison sample
\citep[e.g.,][]{gouldingetal2010}.

Thirdly, even if a black hole is completely inactive, and thus
undetectable by most of the methods discussed here, every once in a
long while a star will wander too close to the event horizon and get
disrupted.  Many likely tidal disruption events have been observed
\citep[e.g.,][]{gezarietal2012,bloometal2011}, likely from stars
falling into $\sim 10^6-10^7$~\msun\ black holes.  Since tidal
disruption events are rare, with at most one every $10^5$ years per
galaxy expected \citep{magorriantremaine1999}, we must monitor many
galaxies every year to detect tidal disruption events.  Ongoing and
upcoming projects are designed to look at the same part of the sky
again and again over years; with these surveys we can hope to detect
many tidal disruptions per year, even around low-mass black holes
\citep{strubbequataert2009}.  Eventually, as the surveys progress, we
may be able to use the detection rate of tidal disruptions in small
galaxies as a indicator of the occupation fraction. There are many
surprises still to come.

\acknowledgements
I thank A. J. Barth, E.~Gallo, L. C. Ho, B. Miller, A. Reines, A. Seth, and M. Volonteri for useful comments.


\end{document}